\documentclass[10pt]{article}
\usepackage[OE]{express}

\begin{document}

\title{Optical readout of hydrogen storage in films of Au and Pd}

\author{Yoshiaki Nishijima,\authormark{1,*} Shogo Shimizu,\authormark{1}, Keisuke Kurihara,\authormark{1} Yoshikazu Hashimoto,\authormark{1} Hajime Takahashi\authormark{2}, Armandas Bal\v{c}ytis,\authormark{3,4} Gediminas Seniutinas,\authormark{3,5} Shinji Okazaki,\authormark{2} Jurga Juodkazyt\.{e},\authormark{4} Takeshi Iwasa,\authormark{6} Tetsuya Taketsugu,\authormark{6} Yoriko Tominaga,\authormark{7} Saulius Juodkazis\authormark{3,8}}
\address{\authormark{1}Department of Physics, Electrical and Computer Engineering, Graduate School of Engineering, Yokohama National University, 79-5 Tokiwadai, Hodogaya-ku, Yokohama 240-8501, Japan.\\
\authormark{2}Department of Materials Science and Engineering, Graduate School of Engineering, Yokohama National University, 79-5 Tokiwadai, Hodogaya-ku, Yokohama 240-8501, Japan.\\
\authormark{3}Centre for Micro-Photonics, Faculty of Science, Engineering and Technology, Swinburne University of Technology, Hawthorn, VIC 3122, Australia\\
\authormark{4}Center for Physical Sciences and Technology, A. Go\v{s}tauto 9, LT-01108 Vilnius, Lithuania.\\
\authormark{5}Paul Scherrer Institute, Villigen CH-5232, Switzerland\\
\authormark{6}Department of chemistry, Faculty of science, Hokkaido University, N10, W8, Kita-ku Sapporo, Hokkaido 060-0810 Japan.\\
\authormark{7}Graduate School of Advanced Sciences of Matter, Hiroshima University, 1-3-1, Kagamiyama, Higashihiroshima, Hiroshima 739-8530, Japan.\\
\authormark{8}Melbourne Center for Nanofabrication, Australian
National Fabrication Facility, Clayton, VIC~3168, Melbourne,
Australia.}

\email{\authormark{*}nishijima-yoshiaki-sp@ynu.ac.jp} 



\begin{abstract}
For hydrogen sensor and storage applications, films of Au and Pd
were \emph{(i)} co-sputtered at different rates or \emph{(ii)}
deposited in a sequential layer-by-layer fashion on a cover glass.
Peculiarities of hydrogen uptake and release were optically
monitored using 1.3~$\mu$m wavelength light. Increase of optical
transmission was observed for hydrogenated Pd-rich films of
10-30~nm thickness. Up to a three times slower hydrogen release
took place as compared with the hydrogen uptake. Compositional
ratio of Au:Pd and thermal treatment of films provided control
over the optical extinction changes and hydrogen uptake/release
time constants. Higher uptake and release rates were observed in
the annealed Au:Pd films as compared to those deposited at room
temperature and were faster for the Au-richer films. Three main
parameters relevant for sensors: sensitivity, selectivity,
stability (reproducibility) are discussed together with the
hydrogenation mechanism in Au:Pd alloys.
\end{abstract}




\section{Introduction}

Plasmonic applications are expanding into a wider spectral range
from UV to THz by harnessing peculiarities of dielectric
permittivity, $\widetilde{\varepsilon}$, in different materials at
the wavelengths of interest. Strong changes of the real and
imaginary parts of the permittivity occur as a result of phase
transitions, phase intermixing or chemical reactions~\cite{Earl}.
For example, the affinity of Pd to hydrogen, which can be absorbed
with an anomalously high volume ratio of $\sim 600$ for solid
Pd~\cite{Corr}, alters the optical response of Pd via changes in
effective permittivity~\cite{Tittl,Wadell,Strohfeldt}. For future
solar hydrogen applications and in fuel
cells~\cite{Nocera,10oe147,Gahleritner,May}, storage and
monitoring of hydrogen is of paramount importance. Hence, sensing
capabilities have to keep pace with the rapid development in
solar-to-hydrogen conversion, which in current state of the art
industrial installation has reached 24.4\%~\cite{Nakamura}.

Detection of hydrogen using non-contact optical readout was
demonstrated by orientational optical plasmonic
scattering~\cite{Djalalian} and spectral shifts in polarization
sensitive extinction~\cite{Tittl,Wadell}. However, a more
sensitive optical detection of Pd hydrogenation can be obtained by
transmission measurements~\cite{Avila}. A "switchable mirror" that
have a rare metal (Gd or Y) mirror with Pd thin film on top shows
a large transmission change in visible wavelength
region~\cite{switch1, switch2}. Such sensors can provide a highly
desirable non-contact method of hydrogen sensing and monitoring,
especially in micro-chip applications where standard electrical
resistance measurement is not desirable due to feasibility of a
dielectric breakdown/discharge~\cite{03jjap4464,03jjap4593}.

Despite the exceptional affinity of Pd towards hydrogen other
important parameters, such as response times and stability should
ideally be fine-tuned for specific applications. The primary way
of controlling the properties of metals is through creation of
alloys. Permittivity control by co-sputtering of Ag and Au has
been demonstrated to provide a possibility to engineer the optical
response of an binary alloy~\cite{Nishijima,marina1,marina2} and
wasrecently extended to ternary Au, Ag and Cu alloys~\cite{16sr}.
Solid solutions of Rh and Ag mixed at the atomic level can store
hydrogen much like Pd, a functionality non-existent either in pure
Rh or Ag\cite{kitagawa2}. Another development of great interest is
that Pd and Ru alloy nano-crystals enable CO oxidization catalysis
superior to that of Rh~\cite{kitagawa1}. Also, Au and Pd alloy
nanoparticle enhances electro catalytic
activities~\cite{torimoto}. Alloys tailored at the nanoscale are
expected to bring forward new functionalities due to particular
geometric configuration of atoms and strain resulting from the
mismatch of lattice constants~\cite{Juarez}, as well as due to
different electron binding and charge distribution which alters
their chemical behavior~\cite{kitagawa3}.

Here, we prepare films of Au and Pd at different alloy intermixing
ratios for prospective applications in hydrogen detection and
storage. Optical readout of transmittance changes during hydrogen
uptake and release was monitored using a simple non contact
method. Thereby the effect of alloy composition as well as
alloying conditions on sensitivity to ambient hydrogen was
elucidated. Detailed understanding of basic reactions with
hydrogen are important for designing optical hydrogen sensor
devices.

\section{Samples and Methods}


\begin{figure}[t]
\begin{center}
\includegraphics[width=11cm]{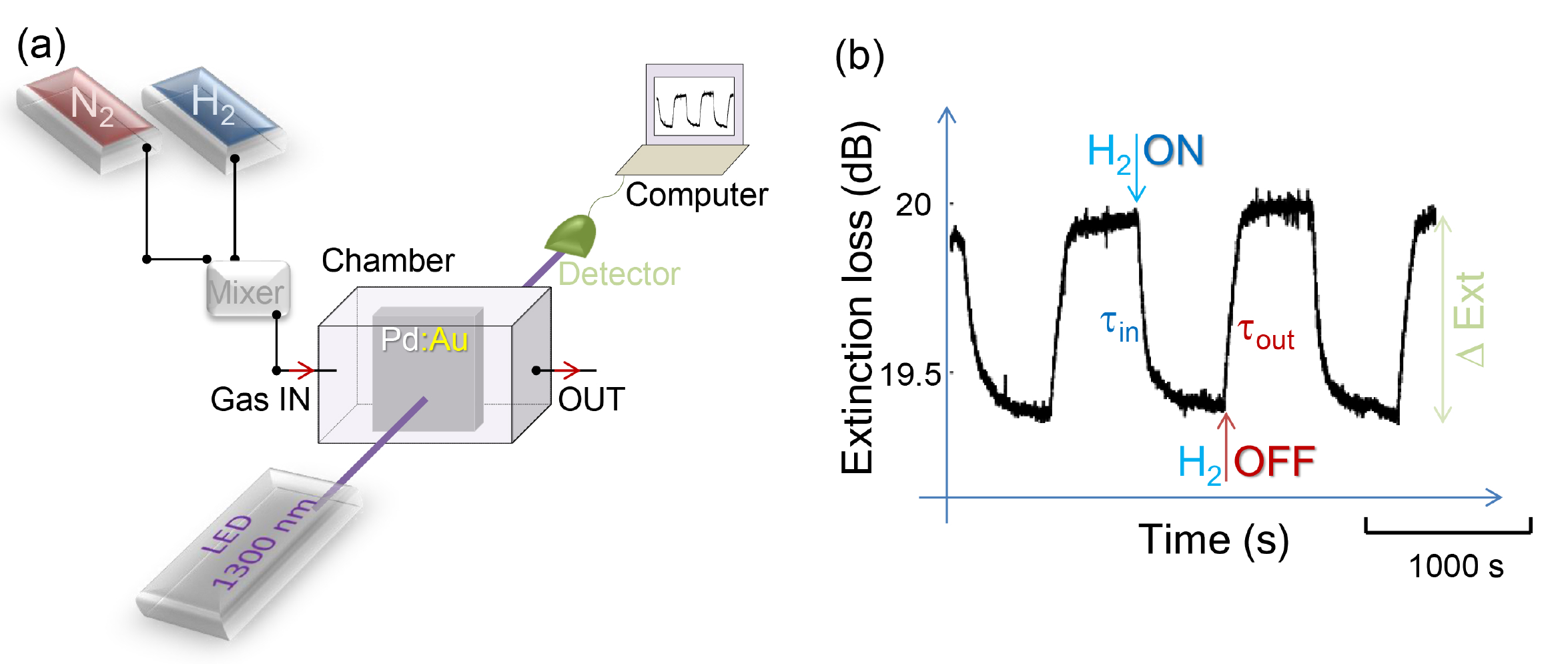}
\caption{(a) Setup for measurement of hydrogenation. (b)
Extinction losses $Ext \equiv -10\lg(I_t/I_0)$~[dB] where
$I_{t,0}$ are the transmitted and reference intensities,
respectively. Hydrogenated Pd becomes more transparent. $\Delta
Ext$ is the change of extinction losses between hydrogen saturated
and depleted states.} \label{f-setup}
\end{center}
\end{figure}

A series of Au and Pd intermixed samples were prepared by
magnetron sputtering (Axxis, JKLesker) using two modes of
deposition: \emph{(i)} co-sputtering (Au:Pd) and \emph{(ii)}
alternating Au and Pd layers (Au-Pd) with different thicknesses
deposited onto a 0.4-mm-thick cover glass. In the latter case, the
numbers of alternation cycles were: 2, 4, 10, 20 (a pair of Au and
Pd layers per one cycle) with corresponding effective thicknesses
of 7.50, 3.75, 1.50, 0.75~nm/cycle, respectively. Sputtering rate
was calibrated using the optical interference microscope 3D
profiler (Bruker Co. Ltd.). Also the thickness of multi-layered
sample were analyzed using X-ray diffraction (XRD). During
sputtering, the substrate was either kept at room temperature or
heated at 250$^\circ$C. At normal conditions, Pd has its native
surface~\cite{Pourbaix}.

The density of states (DOS) for Au, Pd, and AuPd alloys are
calculated at the structures optimized using the
Perdew-Burke-Ernzerhof generalized gradient approximation
(PBE)~\cite{1} and the projector augmented wave method~\cite{2} as
implemented in the VASP code~\cite{3,4,5,6}. The Au-Pd alloys are
constructed by alternating Au and Pd layers in 2-by-2 manner in
the $\langle 001\rangle$ direction using the unit cell consisting
of Au$_4$Pd$_4$ and 1-by-2 in the $\langle 111\rangle$ direction,
using the unit cell consisting of AuPd$_2$. For both types of
alloy systems, a 21 $\times$ 21 $\times$ 11 $k$-point mesh is used
with the cutoff energy of 500~eV.

Optical permittivity of the samples was determined by transmission
and reflection measurements using setup and analysis method
reported earlier~\cite{16sr} at normal incidence. For
hydrogenation of samples, all the optical setup was put into a
sealed container with dimensions 50$\times$50$\times$15 cm$^3$.
The spectral range from 450 to 1000~nm was used as defined by
limits of the used detectors.

\begin{figure}[t]
\begin{center}
\includegraphics[width=11cm]{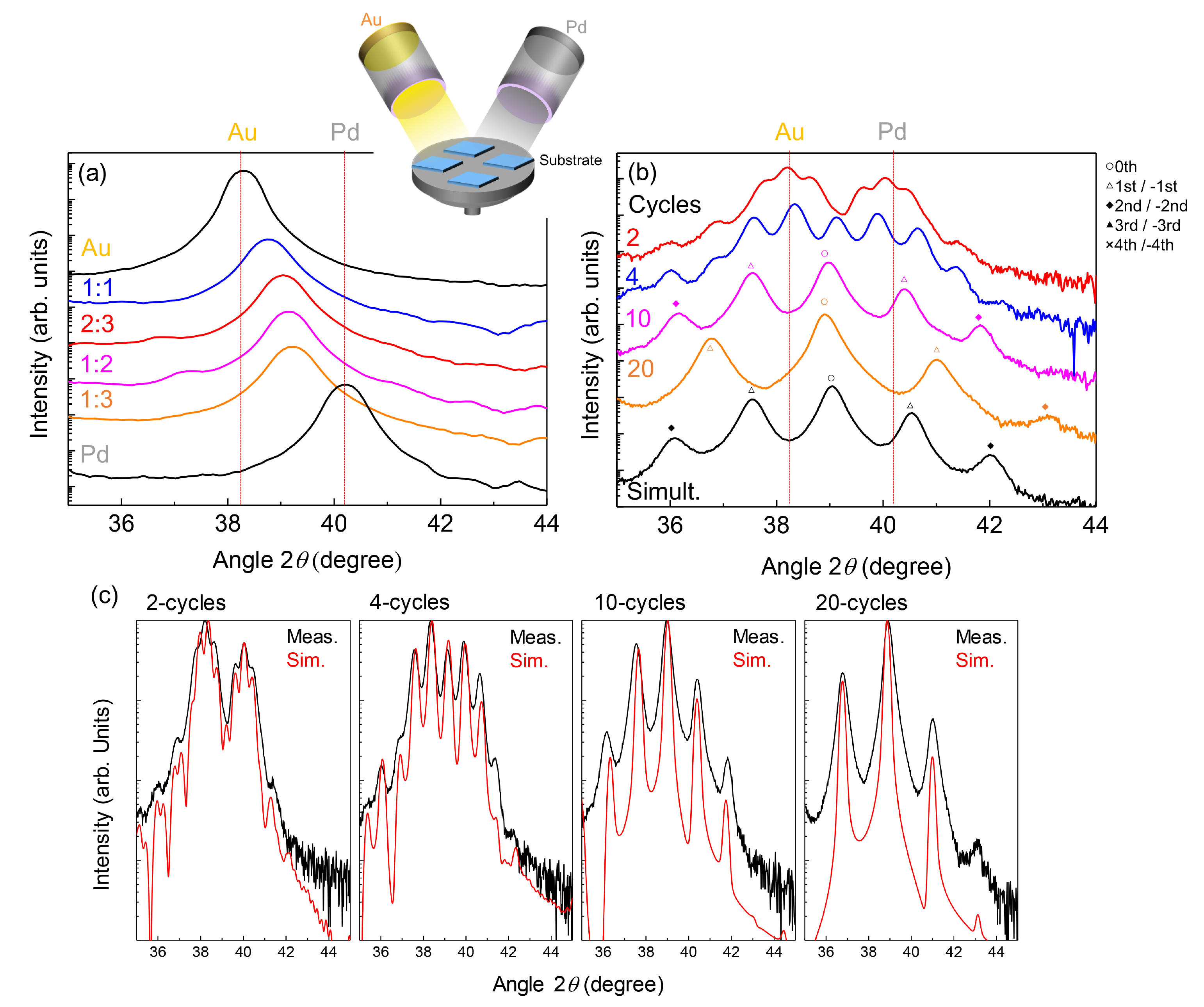}
\caption{(a) XRD of Au:Pd films co-sputtered onto a 250$^\circ$C
pre-heated glass substrate at different sputtering rates.
 Vertical lines mark $\langle111\rangle$ position of the fcc-Au and Pd peaks.
(b) XRD of layered Au:Pd films in the range 40 to 70~nm thickness
deposited at alternating cycles: 2, 4, 10, 20 (corresponding
thicknesses are 7.50, 3.75, 1.50, 0.75~nm, respectively) and
simultaneously deposited. (c) The experimentally measured XRD spectra (Meas.) and simulation (Sim.) results of alternating sputtering } \label{f-xrd}
\end{center}
\end{figure}

Due to the flow rate limitation, the time dependence of hydrogen uptake was measured using a setup (Fig.~\ref{f-setup})
with a 21.4~cm$^3$ sample chamber equipped with a gas mixing
system to provide 500~mL/hour flow rate of a 4\% H$_2$:N$_2$
mixture at a pressure of 0.1~MPa ($\sim 1$~bar), a 1.31~$\mu$m
wavelength laser emitting diode (ILX lightwave Co. MPS- 8012) and
a Ge photo diode detector (Ando Electric Co., Ltd., AQ2150A) was
used to \emph{in situ} monitor transmission changes. A fiber optic
collimator/focuser with a numerical aperture $NA = 0.25$ (Edmund
Optics, Ltd.) was used to define the 2-mm-diameter spot on the
metal film inside the gas chamber~\cite{Okazaki}. The used
wavelength is out of the range used for the permittivity
measurements. However for the most of metals, permittivity can be
approximated by the Drude-Lorenz theory. Therefore behavior at
1.31 $\mu$m can be extrapolated from the results obtained in the
visible stectral range following generic dependence:
\begin{equation}\label{e1}
\varepsilon(\omega) = \varepsilon(\infty) -
\frac{\omega_{p}^{2}}{\omega^2 + i\omega/\tau} + \Sigma_{j}
\frac{A_j \hbar \omega_{0,j}}{(\hbar \omega_{0,j}) - (\hbar
\omega)^2 - i\hbar \omega/\tau_{j}},
\end{equation}
\noindent where the first two terms represent the Drude free
electron model and the last term is the Lorentz contribution
accounting for the bound electrons participating in interband
transitions. Here $\varepsilon(\infty)$ is the permittivity at the
high frequency limit (infinity), $\omega_{p}$ is the plasma
frequency, $\tau$ is the relaxation time of the free electrons;
$j$ denotes the index of a Lorentz oscillator, $A_j$,
$\omega_{0,j}$ and $\tau_{j}$ are the amplitude, the resonant
frequency and the relaxation time of the given oscillator $j$,
respectively,   $\hbar$ is the reduced Plank constant, $i$ is the
imaginary unit, and $\omega$ is the optical cyclic frequency.

\section{Results and discussion}

\subsection{Au:Pd alloy formation during sputtering}
For hydrogen uptake and storage investigations, films of Au and Pd
were prepared on a cover glass by co-sputtering both Au and Pd
metals at different rates or by using layer-by-layer alternating
sputtering. Pd and Au are known to be completely miscible to form
solid solutions~\cite{Juarez,Rusiskaknyga}. Coatings of different
thicknesses were deposited at room temperature (RT) or at
250$^\circ$C substrate temperatures. Structure of the films was
inspected using XRD and showed a continuous intermixing of Au and
Pd in the case of films co-sputtered onto a 250$^\circ$C heated
substrate at different sputtering rates (Fig.~\ref{f-xrd}(a)). A
continuous shift of the $\langle 111\rangle$ peak without a
significant change of its width is indicative of well controlled
alloying of Au and Pd and was better than was obtained by chemical
exchange reactions on colloidal nanoparticles~\cite{Teng}. On the
other hand, for Au-Pd films made by alternating sputtering of Au
and Pd a complex XRD pattern was observed. This reflects a
nano-cluster structure, since the constituent 2-3~nm thickness
films (as judged by sputtering time) do not result in a continuous
fully intermixed material (Fig.~\ref{f-xrd}(b)).

Alloying of Au and  Pd was previously studied as a diffusion
constant, $D$, dependent process~\cite{aupdalloy}. The intermixing
rate was governed by $D$ values of $\sim 10^{-10}$cm$^2/s$ at
1000$^{\circ}$C and $D\sim 10^{-15}$cm$^2$/s at 350$^{\circ}$C. At
the lower temperature of 250$^{\circ}$C, the expected values of
$D$ range from 10$^{-16}$ to 10$^{-17}$cm$^2$/s, and are even more
significantly diminished at RT. Therefore, appreciable migration
of atoms is not expected at RT.

By alternating sputtering, multi layers of Au and Pd were formed
as observed by XRD. The samples made by 2 and 4 cycles of
sputtering showed characteristic separate Au and Pd XRD profiles
which were not forming a periodic layered structure. However, in
the case of 10 and 20 cycles, the pure Au and Pd profiles have
disappeared and satellite peaks in XRD identified formation of
layered structure.

The multiple satellite peaks are well known in the field of
semiconductor super-lattice materials with similar XRD
profiles~\cite{satelite1,satelite2,satelite3,satelite4}. From the
satellite peaks, it is possible to calculate the thickness of each
layer by using the Bragg equation:
\begin{equation}
\Lambda = \frac{m-n}{\sin \theta_m - \sin \theta_n}
\frac{\lambda}{2},
\end{equation}
\noindent where $\lambda = 0.15148$~nm is the characteristic X-ray
wavelength of the Cu $K_\alpha$ line. The suffixes $m$ or $n$
respectively define the $m$-th or $n$-th peak orders at the XRD
spectra. The diffraction multi-peaks were analyzed using a
simulation program LEPTOS, Bruker AXS, which is based on the
dynamical theory of diffraction (see Fig.\ref{f-xrd} (c)). Such analysis, provides
possibility to cross check results of numerical calculations with
XRD experimental data. A single layer thicknesses $\Lambda$ was
deduced to be 20~nm (for 2 cycles), 12~nm (4 cycles), 6.4~nm (10
cycles), and 3.4~nm (20 cycles), respectively. Total thickness of the film has been estimated to
range from 40 to 68~nm.

Similar satellite XRD peaks to those resulting from alternating
deposition were likewise observed for Pd:Au films co-sputtered
onto a RT substrate. The origin of those satellite peaks in the
case of simultaneous co-sputtering needs more detailed analysis.
However, qualitatively they can be explained by the relative
position of the rotating substrate stage in relation to the
sputtering sources. As the substrate plane is tilted with respect
to both of the sources, a gradient of deposition rates develops
along the diameter of the sample~\cite{marina1}. So, as the
substrate rotates during sputtering, periodic oscillations in
alloy composition emerge and, at RT, diffusion proceeds too slowly
to efface them. The films prepared by
sputtering, each with different structural properties, was tested
next for their hydrogen uptake and release capability.

\subsection{Hydrogen storage and optical properties of metals}
\begin{figure}[t]
\begin{center}
\includegraphics[width=9cm]{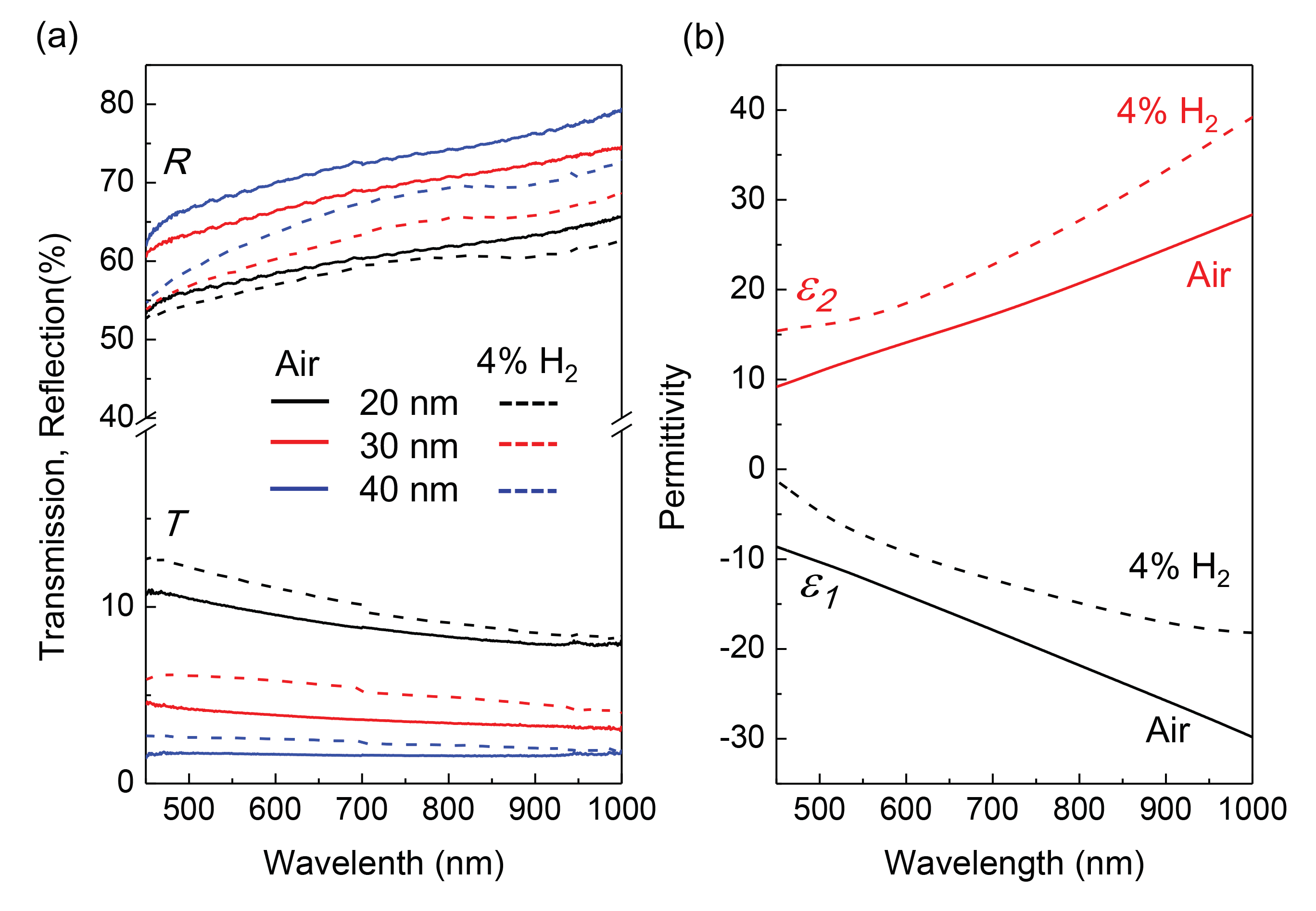}
\caption{(a)Optical transmission and reflection of Pd thin films
(thickness 20, 30 40 nm) with 4\% H$_2$ or N$_2$ (without H$_2$)
condition, and (b) optical permittivity of the Pd with/without
H$_2$} \label{perm}
\end{center}
\end{figure}


Hydrogen storage has to be both reversible and rapid for practical
applications\cite{zhao2}. The interaction between Pd and H affects
their optical properties. Figure~\ref{perm} shows optical
spectroscopic change occurred due to a hydrogen uptake. Typically,
transmittance increased while reflectance decreased. An optical
transmission measurement of a Pd film hydrogenation is sensitive
due to the exponential dependence of transmittance, $T$, on the
imaginary part of the refractive index of a strongly absorbing
film (Pd with hydrogen) $\tilde{n} = n + ik$ on a non-absorbing
substrate of refractive index, $n_s$ (glass)~\cite{Avila}:
\begin{equation}\label{et}
T\simeq \frac{16n_s(n^2 + k^2)}{[(n+n_s)^2 + k^2][(1+n)^2 +k^2]}e^{-\frac{4\pi k d}{\lambda}},\\
R\simeq \frac{(1-n)^2 + k^2}{(1+n)^2 + k^2}
\end{equation}
\noindent where $d$ is the thickness of the Pd film and $\lambda$
is the wavelength of light. With these results, applying the
Drude-Lorenz model, the real and imaginary part of the
permittivity can be determined in the measured wavelength range.
The Drude part, which consists of the $\omega _p$ and $\tau$ give
us  the important information for the free electron behavior.
Where $\omega _p$ is the ratio of free carrier density and
effective mass. $\tau$ is the relaxation time, which indicate the
information of the electron scattering by the electron, hole,
grain boundaries and the other scattering sources. When the
hydrogen was absorbed into the Pd, the $\omega_p$ slightly
increased from 1.5$\times$10$^{16}$ s$^{-1}$ to
1.9$\times$10$^{16}$ s$^{-1}$, $\tau$ decreased from 4.3$\times$
10$^{-16}$ s to 3.0$\times$ 10$^{-16}$~s.
These optical parameter change would be related to the resonance wavelenth shift of the plasmon materials\cite{Nishijima,hydrogen, tittl2}.

This increase of $\omega _p$ means the an increase of the free
electron density and/or decrease of an effective mass of electron.
Hydrogen might donate a free electron to the Pd or softened the
bonding as discussed below. The decreasing $\tau$ is caused by the
increasing of scattering inside the solid phase. The internally
trapped hydrogen causes the scattering of electrons, therefore,
the faster relaxation of free electron oscillation is obtained.

The absorbed hydrogen causes the binding between Pd and hydrogen
tentatively describe as PdH$_x$. Based on thermodynamic arguments
there is a possibility of H$^{+}$ or H$^{+}_2$ species inside the
metal\cite{14ass13,11ass743} Earlier studies showed that upon
hydrogenation of Pd there is heat generation, while for H$_2$
release an elevated temperature of $> 150^\circ$C is
required~\cite{Wei}. It is noteworthy that recent density
functional calculations~\cite{Zhao} of H$_{2}$ interaction with
Au:Pd binary clusters revealed that H$_{2}$ molecules donates
electrons to metal clusters in molecular adsorption. Hydrogen
enters interior of Pd with larger inner surface area. Similar to
longevity of fuel cells, batteries, and super-capacitors,
reversibility of processes at the microscopic level during
charging-discharging are critically important. In this case, the
hydrogen uptake - release.

From the thermoelectric power experiment, hydrogen uptake causes
an increase of electron density in the Pd host. However, the
electrical resistivity is known to increase upon hydrogen uptake
and, at room temperature, can be almost doubled at full
hydrogenation, which depends on pressure~\cite{Holmes,NASA}:
\begin{equation}\label{e0}
\frac{H}{Pd} = 0.69 + \frac{\ln p_H/p_0}{36.8},
\end{equation}
\noindent where $\frac{H}{Pd}$ is the atomic ratio of H and Pd,
$p_H$~[bar] is the hydrogen pressure, and  $p_0$ is the total pressure
of the gas mixture.
The electrons and hydrogen cations in Pd create an interesting guest-host system with an amazingly high
ratio of $\frac{H}{Pd} \simeq 0.95$ at high $16.5\times 10^3$~bar
pressure~\cite{NASA}; $\frac{H}{Pd} = 0.69$ at a hydrogen pressure
equal to 1~bar (at normal conditions).

The results of this study on the optical response of the hydrogen
uptake/release agrees well with electrical response.

For hydrogen release from the solid Pd phase, a dramatic increase
in size occurs from protons to H$_2$ directly or via an
intermediate H$_2^+$. This causes a steric hindrance for hydrogen
release at the surface of Pd and, consequently, it takes longer.
It is noteworthy, that the (bulk$\ominus|\oplus$gas) electron-ion
pair mechanism is fully reversible, hence, is promising for
practical storage applications where reversibility is the key
requirement.


\begin{figure}
\begin{center}
\includegraphics[width=10cm]{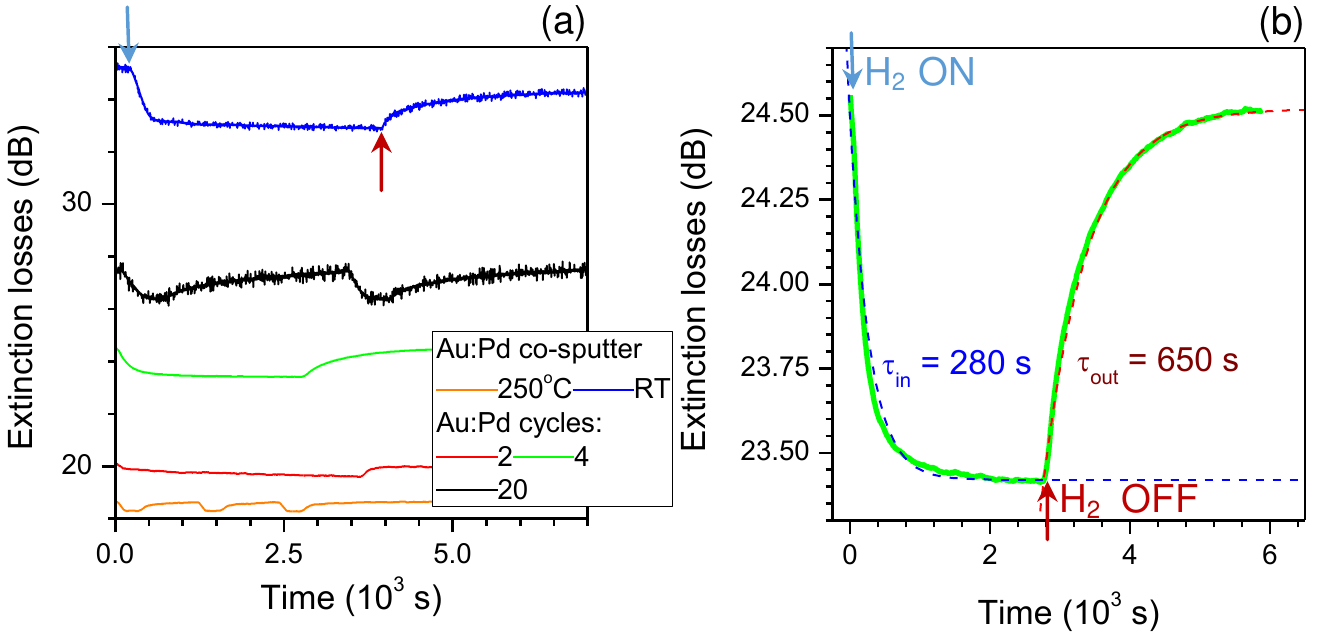}
\caption{Layers of Au-Pd made by alternating deposition. (a)
Extinction [dB] changes during ON/OFF cycling of H$_2$ flow
(marked by arrows) in the case of differently prepared Au-Pd
layers at different mixing ratios deposited on RT or 250$^\circ$C
glass substrates; layers of Au and Pd were sputtered individually
in cycles of 2, 4, 20 with total layer thickness in the range 40
to 70~nm. (b) Transient of the Au-Pd 4-cycle sample fitted with
single exponential uptake and release time constants
$\tau_{in,out}$, respectively. } \label{F-exp}
\end{center}
\end{figure}
\begin{figure}
\begin{center}
\includegraphics[width=9cm]{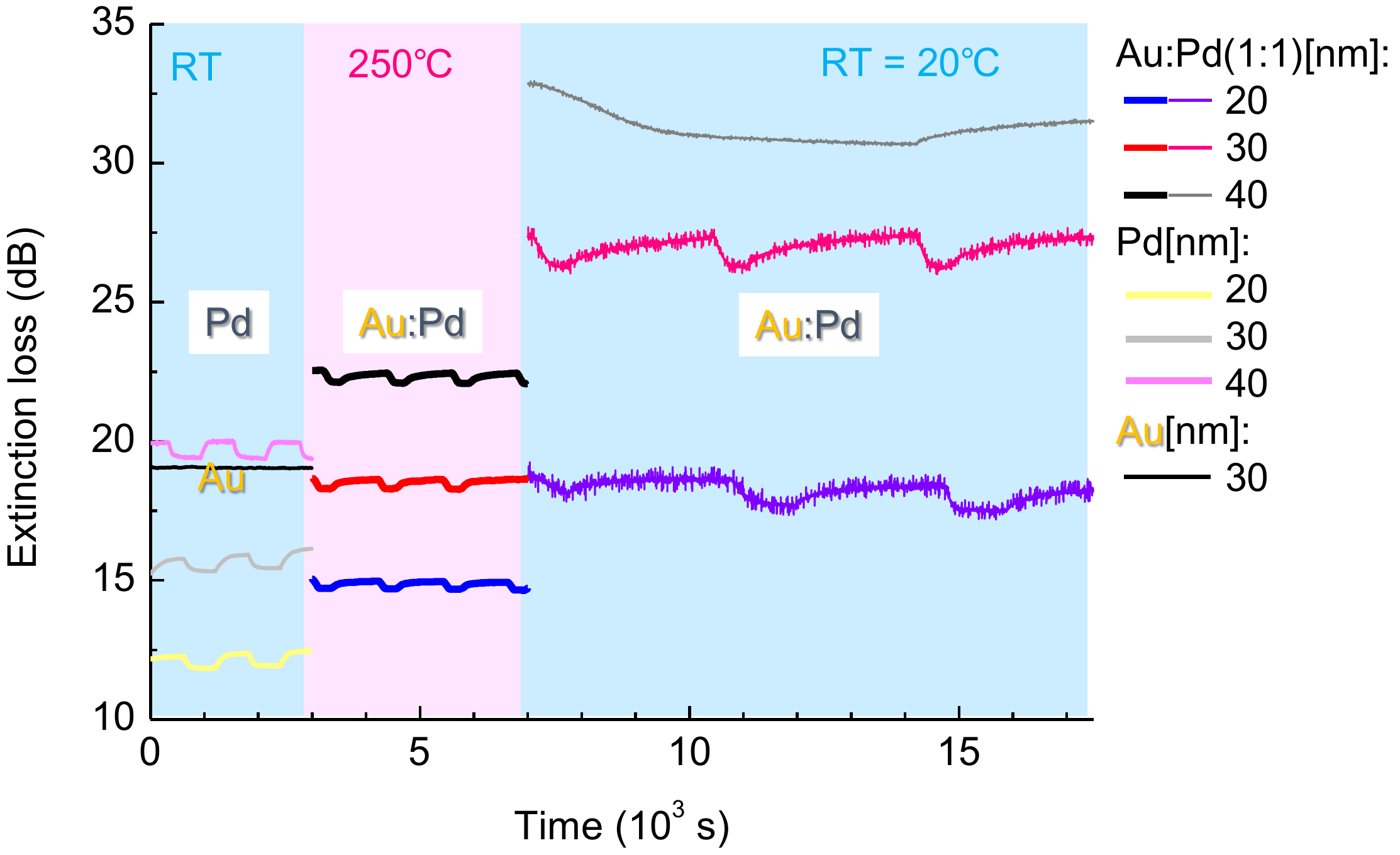}
\caption{Layer of Au:Pd made by co-sputtering. (a) Extinction
losses, $Ext$, (Fig.~\ref{f-setup}(b)) during hydrogenation of
pure Pd and Au:Pd (1:1) co-sputtered films of different thickness
on pre-heated 250$^\circ$C  and RT glass substrates. For Au, $Ext
= Const$ and not dependent on H$_2$ presence (horizontal line for
Au of 20~nm).} \label{f-cycle}
\end{center}
\end{figure}

\subsection*{Optical readout of hydrogenation}

A simple flow chamber with 4$\%$ H$_2$ in a N$_2$ carrier gas was
set up with the possibility to monitor optical transmission
changes upon hydrogenation of Au and Pd films at $\lambda =
1.31~\mu$m wavelength (Fig.~\ref{f-setup}). Extinction losses were
determined by measuring optical transmission: $Ext \equiv
-10\lg(I_t/I_0)$~[dB] where $I_{t,0}$ respectively are the
transmitted and reference intensities; reference intensity $I_0 =
12.2~\mu$W was measured without the sample in a N$_2$ filled
chamber. Following the introduction of H$_2$ no measurable change
in optical transmission was observed for pure Au films with
thickness in the range of 10-30~nm.
\begin{figure}
\begin{center}
\includegraphics[width=10cm]{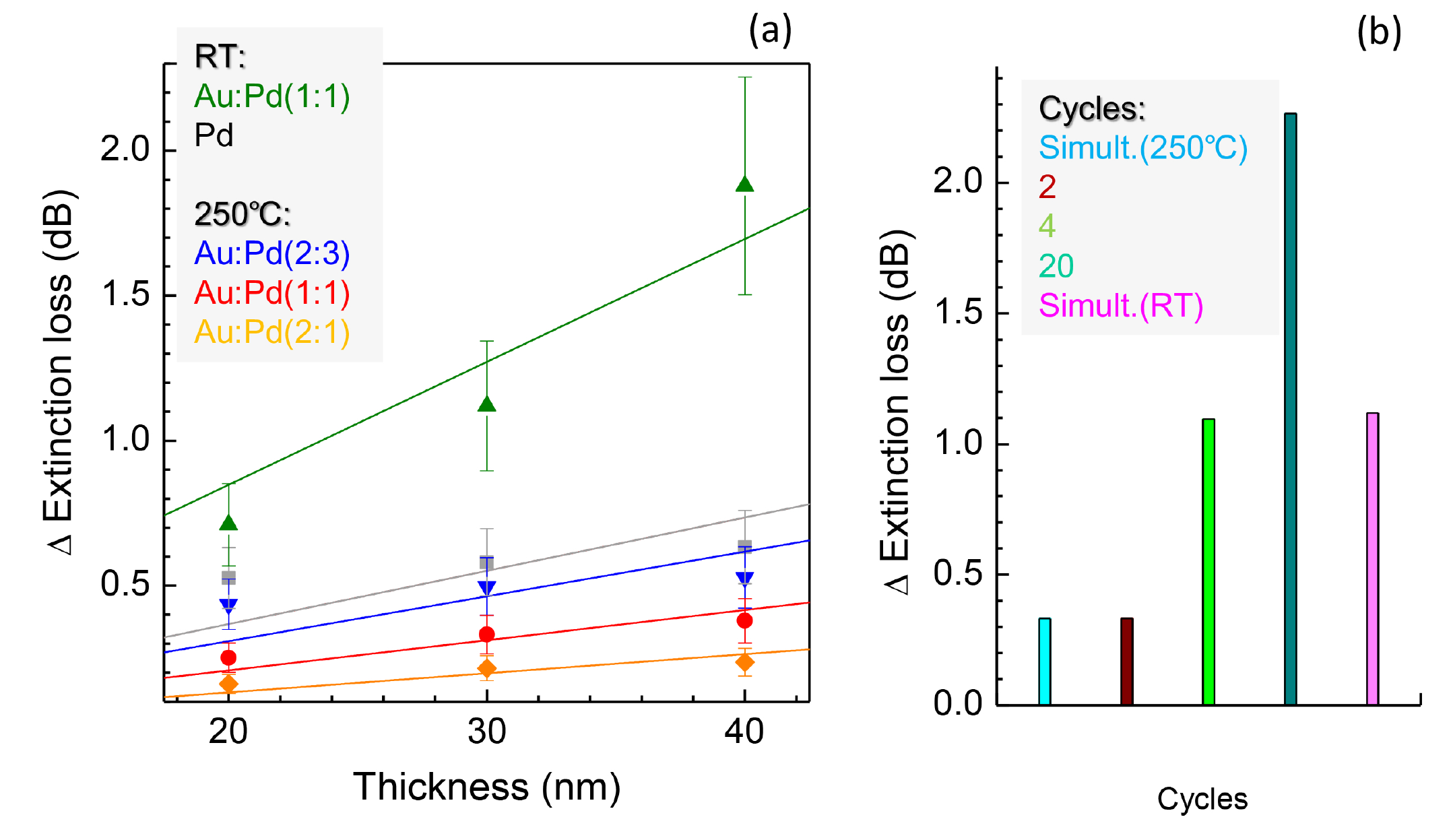}
\caption{(a) Change of extinction, $\Delta Ext$, for
simultaneously co-sputtered films. Error bars are $\pm 20\%$ and
the lines are guides for eye with starting point at origin of
coordinates (0;0). (b) $\Delta Ext$ for Au:Pd (1:1) multi layered
and simultaneous sputtering (250$^\circ$C and RT).}\label{f-ext}
\end{center}
\end{figure}
In the case of Pd containing films the presence of H$_2$ induced
an increase in their transmittance. Figure~\ref{f-cycle} shows the
total extinction of pure Pd as well as alloyed Au:Pd films and
$Ext$ changes as H$_2$ flow was cycled through the chamber. Of
note is that the trend of decreased extinction with progressive
hydrogenation enables increased precision of the measurement due
to a better signal-to-noise ratio as more hydrogen is absorbed by
Pd and a larger change in the extinction, $\Delta Ext$, is
measured. Temporal evolution of hydrogen uptake and release were
well fitted by single exponential transients (Eqn.~\ref{et}) with
time constants $\tau_{in,out}$, respectively
(Fig.~\ref{f-setup}(b)). For the Au:Pd film, there was an obvious
asymmetry $\tau_{in} < \tau_{out}$ (Fig.~\ref{f-cycle}).

Films of Pd and Au:Pd sputtered onto a pre-heated substrate showed
smaller optical losses in transmission, presumably due to a better
homogeneity and smaller nano-porosity, as tested for several
different thicknesses (Fig~\ref{f-cycle}). Figure~\ref{f-ext}
shows the change of extinction losses due to hydrogenation (an
increase in transmission) for differently prepared samples.
Extinction [dB] was linearly dependent on thickness of Pd as one
would expect from optical absorption. Interestingly, mixture Au:Pd
(1:1) sputtered on substrate at RT showed the largest changes of
extinction (Fig.~\ref{f-ext}(a)), even larger than in the case of
pure Pd. When deposition of Pd and Au layers was carried out on a
pre-heated substrate, smaller $\Delta Ext$ values were observed
(Fig.~\ref{f-ext}(b)). This could also be attributed to a greater
extent of alloy lattice ordering in the case of alloy formation on
a pre-heated substrate.The value of $\Delta Ext$ was found to
scale in proportion to the amount of Pd in the intermixed films.
\begin{figure}
\begin{center}
\includegraphics[width=10cm]{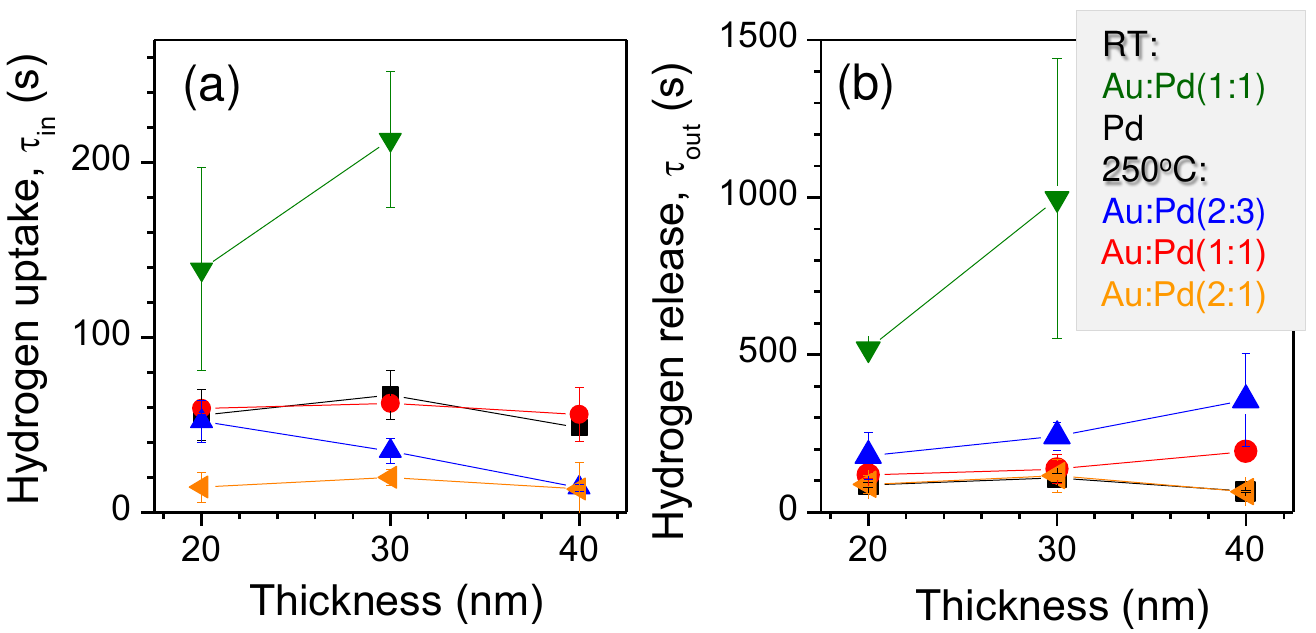}
\caption{Hydrogen uptake (a) and release (b) time constants
$\tau_{in,out}$, respectively, for Au:Pd films prepared by
co-sputtering with and without annealing. } \label{f-time}
\end{center}
\end{figure}

The largest $\Delta Ext$ observed in  Au:Pd (1:1) alloy deposited
on a RT substrate, had the slowest H$_2$ uptake time, $\tau_{in}$
(Fig~\ref{f-time}(a)), markedly slower than in the pure Pd
film~\cite{Teng}. In general, the presence of Au in the film
decreased $\tau_{in}$, most probably, due to a larger electro
negativity of Au (2.54) as compared with Pd (2.22; same as H)
according to the Pauling scale\cite{pauling}. Hydrogen release
time, $\tau_{out}$, was approximately 2-3 times longer compared to
the hydrogen uptake time (Fig~\ref{f-time}(b)). A similar result
was reported in ref.~\cite{Monzon}, where stacked Pd and Au
nano-layers were deposited by thermal evaporation onto optical
fibers and attenuation changes in evanescent wave upon
hydrogenation of a multi-layered film were measured.

\subsection*{Surface adsorption / desorption on Au-Pd alloy}

\begin{figure}
\begin{center}
\includegraphics[width=10cm]{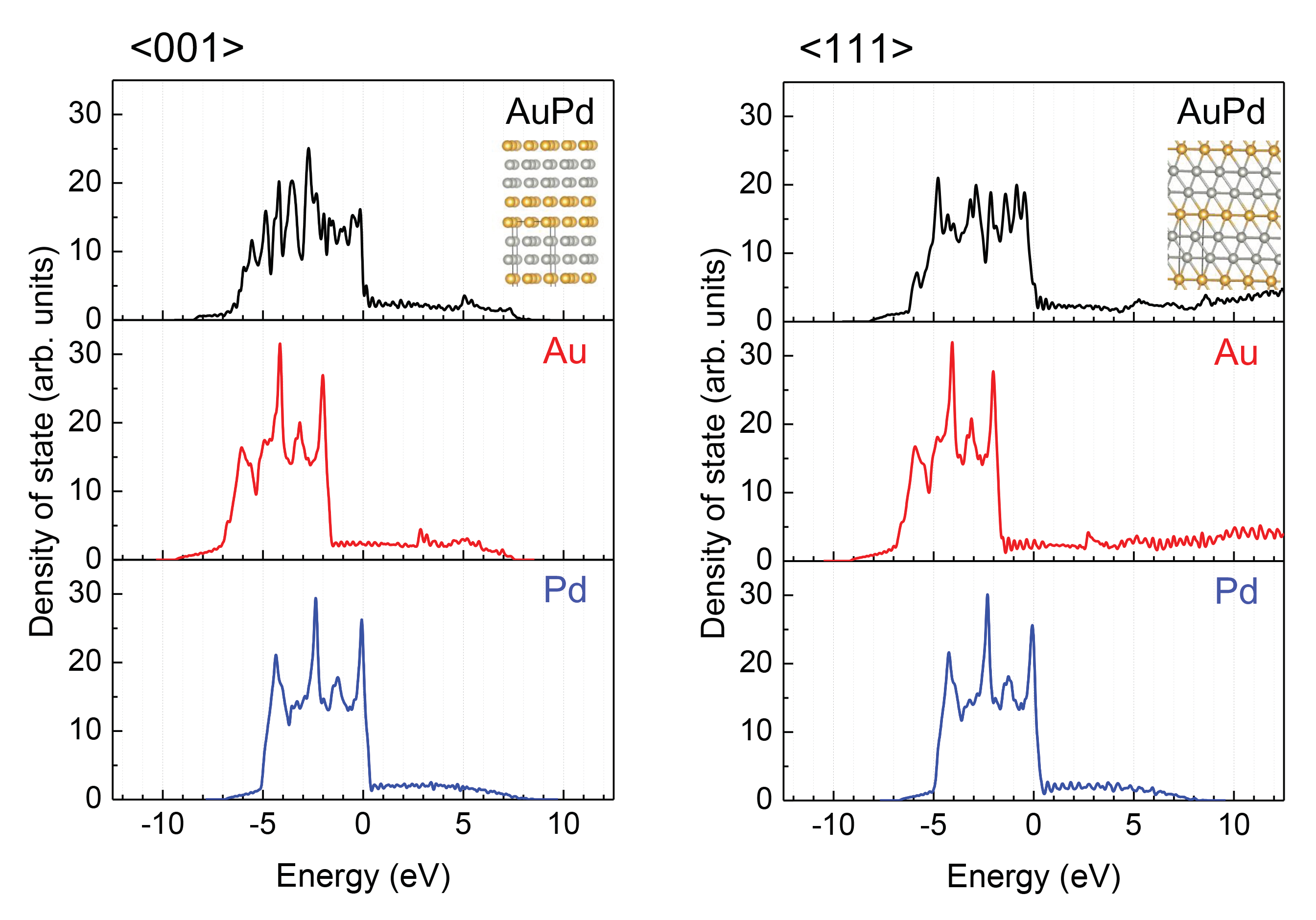}
\caption{The density of state (DOS) for AuPd alloy and pure Au, and Pd systems for the $\langle 001\rangle$ (left column) and $\langle 111\rangle$ (right column) orientations. Gold and silver spheres show Au and Pd, respectively. Zero in the energy x axis indicates the top of the occupied valence band.} \label{SI}
\end{center}
\end{figure}

The hydrogen read out was found to be strongly influenced by Au:Pd
alloying. The hydrogen diffusion constant $D$ in Pd is on the
order of 10$^{-5}$ cm$^2$/s\cite{GLH}. Therefore, the diffusion
time in a thin 10~nm film could be estimated at the order of $\sim
10 \mu$s. This is much more rapid than observed in experiments.
The rate limiting reaction is adsorption/desorption of hydrogen on
the surfaces of metal thin films. Mullins et al. revealed that
temperature defined the hydrogen desorption rate. A lower
temperature desorption has been observed on the surface of Au:Pd
alloying site. This indicated that alloying Au and Pd resulted in
a decrease of affinity to hydrogen~\cite{mullins}. This effect can
be also explained by the density of state (DOS) changes in Au and
Pd alloy. Quaino et al. showed by density functional theory (DFT)
calculations for Au(111) surface on Pd, that the Fermi level of an
alloy was lower in energy due to the Au:Pd bond
formation~\cite{santos}. We have also confirmed this prediction by
DFT calculations in Fig.~\ref{SI}. Hence, an alloy of Au:Pd would
have a reduced activation energy for hydrogen adsorption and
desorption on the surface. Therefore we have obtained faster
uptake and release times in Au:Pd alloy films especially with
annealed and homogeneously alloyed samples.

\section*{Conclusions}

The mixture of Au and Pd made by co-sputtering can improve
(shorten) the response time during H$_2$ uptake and release
stages, most probably, due to electro negativity of Au which
facilitates the adsorption of positively charged hydrogen species
H$_2^+$ and H$^{+}$ at the inner surface of the alloy phase.
Multi-layered sample made by alternating sputtering has higher
sensitivity as compared with co-sputtered samples. Stability in
H$_2$ uptake-release performance showed no difference on a
day-to-day basis. This is consistent with earlier observations
that embitterment during hydrogenation can be avoided~\cite{Wei}.
The fully reversible mechanism of hydrogen uptake into Pd phase is
promising for applications where long term stability is of
paramount importance. Using Au:Pd mixtures it is possible to
enable more rapid and sensitive optical hydrogen monitoring device
construction.

\section*{Acknowledgement}
\small{YN grateful for partial support by Japan Society for the
Promotion of Science (JSPS), Grants-in-Aid for Scientific
Research,  Open Partnership Joint Projects of JSPS Bilateral Joint
Research Projects and Tateishi fundation. Amada fundation, SJ
acknowledges a partial support via the Australian Research Council
Discovery project DP130101205 and a startup funding of
Nanotechnology facility by Swinburne University. The computation
in this work has partly been done using the facilities of the
Supercomputer Center, the Institute for Solid State Physics, the
University of Tokyo. Authors greatly thank to Dr. Hitoshi Morioka
of Bruker AXS K.K., and Professor Masahiro Yoshimoto and Assistant
Professor Hiroyuki Nishinaka of Kyoto Institute of Technology for
their helps with the simulation analysis of XRD spectra.}


\begin{thebibliography}{99}
\bibitem{Earl}S. K. Earl, T. D. James, T. J. Davis, J. C. McCallum, R. E. Marvel, R. F. Haglund, and A. Roberts"Tunable optical antennas enabled by the phase transition in vanadium dioxide," Opt. Express {\bf21}, 27503-27508 (2013).
\bibitem{Corr}A. Baumel, P. Drodten, E. Heitz, and R. Bender, Chap. A 35. Platinum metals (Ir, Os, Pd, Rh, Ru) (John Wiley and Sons, 2008).
\bibitem{Tittl} A. Tittl, C. Kremers, J. Dorfmuller, D. N. Chigrin, and H. Giessen, "Spectral shifts in optical nanoantenna-enhanced hydrogen sensors," Opt. Mat. Express {\bf 2}, 111-118 (2012).
\bibitem{Wadell} C. Wadell, and C. Langhammer, "Drift-corrected nanoplasmonic hydrogen sensing by polarization," Nanoscale {\bf 7}, 10963-10969 (2015).
\bibitem{Strohfeldt} N. Strohfeldt, J. Zhao, A. Tittl, and H. Giessen, "Sensitivity engineering in direct contact palladium-gold nano-sandwich hydrogen sensors," Opt. Mat. Express {\bf 5}, 2525-2535 (2015).
\bibitem{Nocera} N. S. Lewis, and D. G. Nocera, "Powering the planet: Chemical challenges in solar energy utilization," Proc. Nat. Acad. Sci. {\bf103}, 15729-15735 (2006).
\bibitem{10oe147} K. Juodkazis, J. Juodkazyte, P. Kalinauskas, E. Jelmakas, and S. Juodkazis, "Photoelectrolysis of water: Solar hydrogen-achievements and perspectives," Opt. Express: energy express {\bf18}, A147-A160 (2010).
\bibitem{Gahleritner} G. Gahleitner, "Hydrogen from renewable electricity: An international review of power-to-gas pilot plants for stationary applications," Int. J. Hydrogen Energy 38, 2039-2061 (2013).
\bibitem{May} M. M. May, H. J. Lewerenz, D. Lackner, F. Dimroth, and T. Hannappel, "Efficient direct solar-to-hydrogen conversion by in situ interface transformation of a tandem structure," Nature Commun. {\bf6}, 8286 1-7 (2015).
\bibitem{Nakamura} A. Nakamura, Y. Ota, K. Koike, Y. Hidaka, K. Nishioka, M. Sugiyama, K. Fujii, "A 24.4 solar to hydrogen energy conversion efficiency by combining concentrator photovoltaic modules and electrochemical cells," Appl. Phys. Express {\bf8}, 107101 (2015).
\bibitem{Djalalian} A. D. Assl, D. E. Gomez, A. Roberts, and T. J. Davis, "Frequency-dependent optical steering from subwavelength plasmonic structures," Opt. Lett. 37, 4206-4208 (2012).
\bibitem{Avila} J. I. Avila, R. J. Matelon, R. Trabol, M. Favre, D. Lederman, U. G. Volkmann, A. L. Cabrera, "Optical properties of Pd thin films exposed to hydrogen studied by transmittance and reflectance spectroscopy," J. Appl. Phys. {\bf107}, 023504 (2010).
\bibitem{switch1} I. Aruna, B. R. Metha, and L. K. Malhotra, "Faster H recovery in Pd nanoparticle layer based Gd switchable mirrors: Size-induced geometric and electronic effects," Appl. Phys. Lett. 87, 103101 1-3 (2005).
\bibitem{switch2}  I. Aruna, B. R. Mehta, L. K. Malhotra, and S. M. Shivaprasad, "A color-neutral. Gd nanoparticle switchable mirror with improved optical contrast and response time," Adv. Mater. 16, 169-173 (2004).
\bibitem{03jjap4464} S. Tomonari, H. Yoshida, M. Kamakura, K. Yoshida, K. Kawahito, M. Saitoh, H. Kawada, S. Juodkazis, H. Misawa, "Efficient microvalve driven by a Si-Ni bimorph," Jpn. J. Appl. Phys. {\bf42}, 4464-4468 (2003).
\bibitem{03jjap4593} S. Tomonari, H. Yoshida, M. Kamakura, K. Yoshida, K. Kawahito, M. Saitoh, H. Kawada, S. Juodkazis, H. Misawa, "Miniaturization of a thermally driven Ni-Si bimorph. Jpn. J. Appl. Phys. {\bf42}, 4593-4597 (2003).
\bibitem{Nishijima} Y. Nishijima, and S. Akiyama, "Unusual optical properties of the Au/Ag alloy at the matching mole fraction," Opt. Mat. Express {\bf2}, 1226-1235 (2012).
\bibitem{marina1}  C. Gong, and M. S. Leite, "Noble metal alloys for plasmonics," ACS Photon. {\bf3}, 507-513 (2016).
\bibitem{marina2} C. Gong, M. Rebello, S. Dias, G. C. Wessler, J. A. Taillon, L. G. Salamanca-Riba, M. S. Leite, "Near-field optical properties of fully alloyed noble metal nanoparticles," Adv. Opt. Mater. {\bf5}, 1600568, 1-6 (2017).
\bibitem{16sr} Y. Hashimoto, G. Seniutinas, A. Balcytis, S. Juodkazis, and Y. Nishijima, "Au-Ag-Cu nano-alloys: tailoring of permittivity," Sci. Reports {\bf6}, 25010 1-9 (2016).
\bibitem{kitagawa2} K. Kusada, M. Yamauchi, H. Kobayashi, H. Kitagawa, Y. Kubota, "Solid solution alloy nanoparticles of immiscible Pd and Ru elements neighboring on Rh: Changeover of the thermodynamic behavior for hydrogen storage and enhanced CO-oxidizing ability," J. Am. Chem. Soc. {\bf136}, 1864-1871 (2014).
\bibitem{kitagawa1} K. Kusada, M. Yamauchi, H. Kobayashi, H. Kitagawa, and Y. Kubota, "Hydrogen-storage properties of solid-solution alloys of immiscible neighboring elements with Pd," J. Am. Chem. Soc. {\bf132}, 15896-15898 (2010).
\bibitem{torimoto} M. Hirano, K. Enokida, K. Okazaki, S. Kuwabata, H. Yoshida and T. Torimoto "Compositio-dependent electrocatalytic activity of AuPd alloy nanoparticles prepared via simultaneous sputter deposition into an ionic liquid," Phys. Chem. Chem. Phys. {\bf15}, 7286-7294 (2013).
\bibitem{Juarez} M. F. Juarez, G. Soldano, H. Guesimi, F. Tielens, and E. Santos, "Catalytic properties of au electrodes modified by an uderlayer of Pd," Surf. Sci. {\bf631}, 235-247 (2015).
\bibitem{kitagawa3} H. Kobayashi, K. Kusada, and H. Kitagawa, "Creation of novel solid-solution alloy nanoparticles on the basis of density-ofstates engineering by interelement fusion," ACC. Chem. Res. {\bf48}, 1551-1559, (2015).
\bibitem{Pourbaix} M. Pourbaix, "Atlas of electrochemical equilibria in aqueous solutions" National Association of Corrosion Engineers, 1974.
\bibitem{1} J. P. Perdew, K. Burke, and M. Ernzerhof, "Generalized gradient approximation made simple," Phys. Rev. Lett. {\bf77}, 3685, (1996).
\bibitem{2} P. E. Blochl, "Ab initio molecular dynamics for liquid metals," Phys. Rev. B {\bf50}, 17953 (1994).
\bibitem{3} G. Kresse, and J. Hafner, "Ab initio molecular dynamics for liquid metals," Phys. Rev. B {\bf47}, RC558 (1993).
\bibitem{4} G. Kresse, "Ab initio Molekular Dynamik fur flussige Metalle," Ph.D. thesis, Diss., Techn. Universitat Wien (1993).
\bibitem{5} G. Kresse, and J. Furthmuller, "Efficiency of ab-initio total energy calculations for metals and semiconductors using a plane-wave basis set," Comput. Mat. Sci. {\bf6}, 1-50 (1996).
\bibitem{6} G. Kresse, and J. Furthmuller, "Efficient iterative schemes for ab initio total-energy calculations using a plane-wave basis set," Phys. Rev. B {\bf54}, 11169 (1996).
\bibitem{Okazaki} S. Okazaki, and S. Johjima, "Temperature dependence and degradation of gasochromic responsebehavior in hydrogen sensing with Pt/WO3 thin film," Thin Solid Films {\bf558}, 411-415 (2014).
\bibitem{Rusiskaknyga}  A. E. Bon and I. K. Kagan, "Stroyeniye i svoistva dvoinykh metalicheskikh sistem (in Russian)" (Nauka, Moscow, 1976).
\bibitem{Teng} X. Teng, Q. Wang, P. Liu, W. Han, A. I. Frenkel, W. Wen, N. Marinkovic, J. C. Hanson, J. A. Rodriguez, "Formation of Pd/Au nanostructures from pd nanowires via galvanic replacement reaction," J. Am. Chem. Soc. {\bf130}, 1093-1101 (2008).
\bibitem{aupdalloy} M. Murakami, D. deFontaine, and J. Fodor, "X-ray diffraction study of interdiffusion in bimetallic Au/Pd thin films," J. Appl. Phys. {\bf47}, 2850-2856 (1976).
\bibitem{satelite1} H. Nakajima, H. Fujimori, and M. Koiwa, "Interdiffusion and structural relaxation in Mo/Si multilayer films," J. Appl. Phys., {\bf63}, 1046-1051 (1988).
\bibitem{satelite2} I. K. Schuller, "New class of layered materials," Phys. Rev. Lett. {\bf44}, 1597-1600 (1980).
\bibitem{satelite3} B. M. Clemens, and J. G. Gay, "Effect of layer-thickness fluctions on superlattice diffraction," Phys. Rev. B {\bf35}, 9337-9340, (1987).
\bibitem{satelite4} Y. Tominaga, Y. Kinoshita, K. Oe, and M. Yoshimotoa, "Structural investigation of GaAs1-xBix/GaAsGaAs1-xBix/GaAs multiquantum wells," Appl. Phys. Lett. {\bf93}, 131915 (2008).
\bibitem{zhao2} Z. Zhao, M. Carpenter, H. Xi, and D. Welch, "All-optical hydrogen sensor based on a high alloy content palladium thin film," Sensors and Actuators B {\bf113}, 532-538 (2006).
\bibitem{hydrogen} Y. Nishijima, A.  Bal\v{c}ytis, G. Seniutinas, S. Juodkazis, T. Arakawa, S. Okazaki, Raimondas Petru\v{s̆}kevi\v{c̆}ius, "Plasmonic Hydrogen Sensor at Infrared Wavelength," submitted, (2017).
\bibitem{tittl2} A. Tittl, P. Mai, R. Taubert, D. Dregely, N. Liu, and H. Giessen, "Palladium-based plasmonic perfect absorber in the visible wavelength range and its application to hydrogen sensing," Nano Lett., {\bf 11}, 4366-4369, (2011).
\bibitem{14ass13} K. Juodkazis, J. Juodkazyte, B. Sebeka, and S. Juodkazis, "Reversible hydrogen evolution and oxidation on Pt electrode mediated by molecular ion," Appl. Surf. Sci. {\bf290}, 13-17 (2014).
\bibitem{11ass743} K. Juodkazis, J. Juodkazyte, A. Grigucevicien and S. Juodkazis, "Hydrogen species within the metals: role of molecular hydrogen ion H+2," Appl. Surf. Sci. {\bf258}, 743-747 (2011).
\bibitem{Wei} I. Wei, and J. Brewer, "Desorption of hydrogen from palladium plating, AMP J. Technol. {\bf5}, 49-53 (1996).
\bibitem{Zhao} S. Zhao, Z. Z. Tian, J. N. Liu, Y. L. Ren, and J. J. Wang, "Interaction of H2 with gold-palladium binary clusters: Molecular and dissociative adsorption," Computat. Theor. Chem. {\bf1055}, 1-7 (2015).
\bibitem{Holmes} R. M. Holmes, "The effect of absorbed hydrogen on the thermoelectric properties of palladium," Science {\bf LVI}, 201-202 (1922).
\bibitem{NASA} D. A. Otterson, and R. J. Smith, "Absorption of hydrogen by palladium and electrical resistivity up to H-Pd atom ratios of 0.97," NASA report, National Aeronautics and Space Administration, Washington DC (1969).
\bibitem{pauling} L. Pauling, "The nature of the chemical bond. iv. the energy of single bonds and the relative electronegativity of atoms," J. Am. Chem. Soc. {\bf54}, 3570-3582 (1932).
\bibitem{Monzon} D. Monzon-Hernandez, D. Luna-Moreno, and D. Martinez-Escobar, D. "Fast response fiber optic hydrogen sensor based on palladium and gold," Sens. Actuators B: Chem. {\bf136}, 562-566 (2009).
\bibitem{GLH} L. Holleck, "Diffusion and solubility of hydrogen in palladium and palladium-silver alloys," J. Phys. Chem. 74, 503-511, (1970).
\bibitem{mullins} W. Y. Yu, G. M. Mullen, and C. B. Mullins, "Hydrogen adsorption and absorption with Pd-Au bimetallic surfaces," J. Phys. Chem. C {\bf117}, 19535-19543 (2013).
\bibitem{santos} P. M. Quaino, R. Nazmutdinov, L. F. Peiretti, and E. Santos, "Unravelling the hydrogen absorption process in Pd overlayers on a Au(111) surface," Phys. Chem. Chem. Phys. {\bf18}, 3659-3668 (2016).
\end{thebibliography}
\end{document}